\documentclass[%
 reprint,
showpacs,preprintnumbers,showkeys,
 amsmath,amssymb,
 aps,
 prl,
 floats,
floatfix,
]{revtex4-1}

\usepackage{graphicx}
\usepackage{dcolumn}
\usepackage{bm}

\begin{document}

\preprint{MMM2011/CU-10}

\title{Field-induced Magnetic Transition in Cobalt-Ferrite}

\author{Martin Kriegisch}
\email{kriegisch@ifp.tuwien.ac.at}
\affiliation{
Institute of Solid State Physics, Vienna University of Technology\\
Wiedner Hauptstrasse 8-10/E138, A-1040 Vienna, Austria}

\author{Weijun Ren}
\affiliation{
Shenyang National Laboratory for Materials Science, Institute of Metal Research, and International Center for Materials Physics,
Chinese Academy of Sciences, Shenyang 110016, People's Republic of China}

\author{Reiko Sato-Turtelli}
\affiliation{
Institute of Solid State Physics, Vienna University of Technology\\
Wiedner Hauptstrasse 8-10/E138, A-1040 Vienna, Austria}
\author{Herbert M\"uller}
\affiliation{
Institute of Solid State Physics, Vienna University of Technology\\
Wiedner Hauptstrasse 8-10/E138, A-1040 Vienna, Austria}
\author{Roland Gr\"ossinger}
\affiliation{
Institute of Solid State Physics, Vienna University of Technology\\
Wiedner Hauptstrasse 8-10/E138, A-1040 Vienna, Austria}

\author{Zhidong Zhang}
\affiliation{
Shenyang National Laboratory for Materials Science, Institute of Metal Research, and International Center for Materials Physics,
Chinese Academy of Sciences, Shenyang 110016, People's Republic of China}

\date{\today}

\begin{abstract}
We present magnetostriction and magnetization measurements of a cobalt ferrite (Co$_{0.8}$Fe$_{2.2}$O$_{4}$) single crystal. We observe unusual behaviour in the magnetic hard axis of the single crystal which manifests in a jump of the magnetization curve at a critical field. This first order magnetization process (FOMP) which is explained as an anisotropy driven transition  is visible at temperatures lower than 150~K. By analyzing the anisotropy constants we found that the higher order anisotropy constant $K_{2}$ dominates the anisotropy energy. In the magnetostriction measurements the FOMP is clearly visible as a huge jump in the [111] direction, which can be explained by means of a geometric model.
\end{abstract}

%
\maketitle


\section{Introduction}
Cobalt Ferrite is under examination since more than sixty years, but still there is quite a lot of open problems. In 1988 Guillot observed a jump in the field dependence of the magnetization in pure Cobalt Ferrite with the composition Co$_{1.04}$Fe$_{1.96}$O$_{4}$ and also in Cd substituted Cobalt Ferrite \cite{guillot_high_1988}. This jump was explained as a spin-flip. Because of the rather low critical field this explanation was not conclusive. Therefore within the present work this transition was studied in more detail.
\section{Sample Preparation}
The single crystal with the composition Co$_{0.8}$Fe$_{2.2}$O$_{4}$ was grown by flux method. The starting materials are 18~g Na$_{2}$B$_{4}$O$_{7}$~$\cdot$~10H$_{2}$O (Borax), 2.3~g CoO (99.99\%), and 6.7~g Fe$_{2}$O$_{3}$ (99.99\%). After sufficiently mixing, the materials were put in a tightly closed Pt crucible and heated from room temperature to 1370~$^{\circ}$C at a rate of 100~$^{\circ}$C/h, and then held for a period of 6~h; slowly cooled from 1370 to 990~$^{\circ}$C at 2~$^{\circ}$C/h, followed by a furnace cooling by switching off the power supply \cite{wang_flux_2006}. The composition of the single crystal was checked by XRD and SEM investigation.
\section{Experimental Procedures}
Magnetization was measured from 5 to 400~K in a vibrating sample magnetometer with a superconducting 9~T coil. The single crystal was oriented by XRD in a Laue setup and then transfered to the VSM sample holder. The error from transferring the single crystal from one sample holder to the other was usually smaller than 1.5$^{\circ}$.\\
The magnetostriction was measured with a miniature capacitive dilatometer described in \cite{rotter_miniature_1998} using a cryostat with a variable temperature insert (VTI) and a superconducting 9~T coil.\\
\section{Results and Discussion}
\subsection{Magnetization}
\begin{figure}
\includegraphics{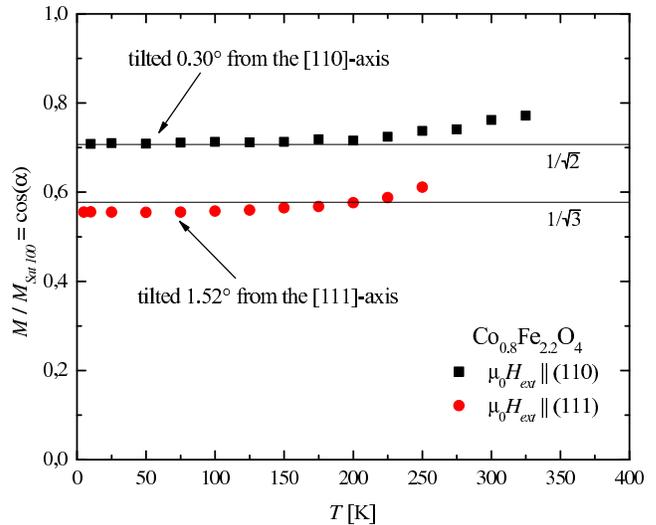}
\caption{\label{fig:Geo}Offset of the linear fits of normalized magnetization of Co$_{0.8}$Fe$_{2.2}$O$_{4}$  as a function of temperature. The straight lines indicate the geometrical fit}
\end{figure}

The degree of inversion of the cation distribution of the inverse spinel (A$^{2+}$)[B$_{2}^{3+}$]O$_{4}$ was calculated to $i = 0.625$ with (Co$_{0.8-i}^{2+}$ Fe$_{0.2+i}^{3+}$)[Co$_{i}^{2+}$Fe$_{2-i}^{3+}$]O$_{4}$ and the magnetic moment of $\mu = \mu_{B-Sites} - \mu_{A-Sites} = 4.1 \mu_{B}$ at $T = 5$~K. The value of saturation magnetization at 5 and 10~K is practically the same due to the the very high ordering temperature.\\
\begin{figure}
\includegraphics{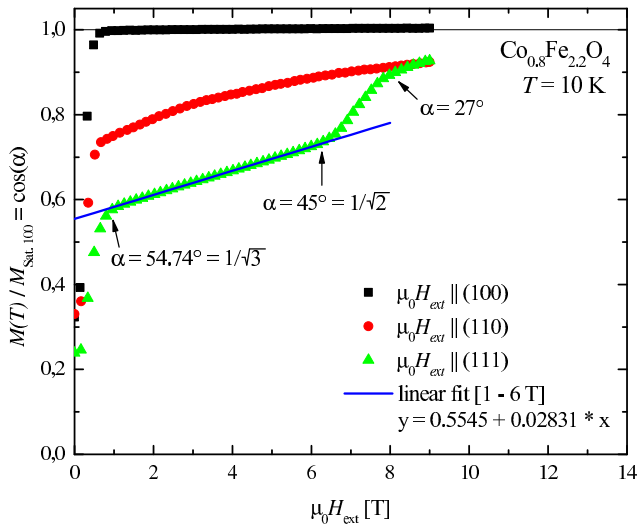}
\caption{\label{fig:MH10K}Normalised Magnetization of Co$_{0.8}$Fe$_{2.2}$O$_{4}$ single crystal at $T = 10$~K}
\end{figure}
The magnetization measurements revealed that the [100] axis is the easy axis of magnetization of Co-ferrite over the whole temperature range. Below 150~K a jump in the magnetization at fields occurs as it was also found in a similar material (Co$_{1.04}$Fe$_{1.96}$O$_{4}$) in Ref.~\cite{guillot_high_1988}. In figure~\ref{fig:MH10K} the normalized magnetization at $T = 10$~K is plotted versus the external magnetic field. The jump is clearly visible at roughly 7~T in the [111] axis. The critical field differs from that published by Guillot which can be understood regarding the different sample composition. A linear fit between 1 and 6~T was performed showing that the extrapolation to 0~T leads to a value of $\frac{1}{\sqrt[]{3}}$, which indicates that the magnetization vector lies in the [100] axis and only the projection ($cos(\alpha)$) of the [100] magnetization vector into the [111] axis is measured. By increasing the magnetic field the vector starts to rotate towards the [111] axis. However as soon as the measured moment in the [111] axis achieves the value of $\frac{1}{\sqrt[]{2}}$, the measured moment becomes equal to the [110] value measured at 0~T. At this point the system needs more energy (field) to overcome the magnetic anisotropy energy corresponding to the [110] axis. As a consequence the magnetization vector rotates in the [110] axis to a local energy minimum in order to reduce energy. In figure~\ref{fig:Geo} the offset of the linear fits assuming such a geometrical model below the transition are shown. The error due to misalignment is found to be 1.5$^{\circ}$ for the [111] axis and 0.3$^{\circ}$ for the [110] axis. Obviously our geometric model works well from 5~K up to 250~K. Above 250~K the anisotropy along the [110] and [111] axis decreases and the magnetization vector can rotate smoothly (without a discontinuity) out of the easy axis.\\
By this geometrical considerations one can see that the origin of the magnetization jump is not a spin-flip, but a rotation in the magnetization due to the higher order anisotropy constants which cause a more complicated shape of the anisotropy energy $E_{A}$.\\
Such anisotropy driven magnetization jumps are generally called FOMP (First Order Magnetization Process) and were also found in other complex systems and alloys (as e.g. PrCo5 - see \cite{asti_abstract:_1979}, RE$_{2}$Fe$_{17}$C \cite{grossinger_anisotropy_1991}, REFe$_{11}$Ti \cite{kou_magnetic_1993} or Nd$_{2}$Fe$_{14}$B \cite{kou_spin-reorientation_1997}).
\subsection{Magnetic Anisotropy}
\begin{figure}
\includegraphics[width=0.47\textwidth]{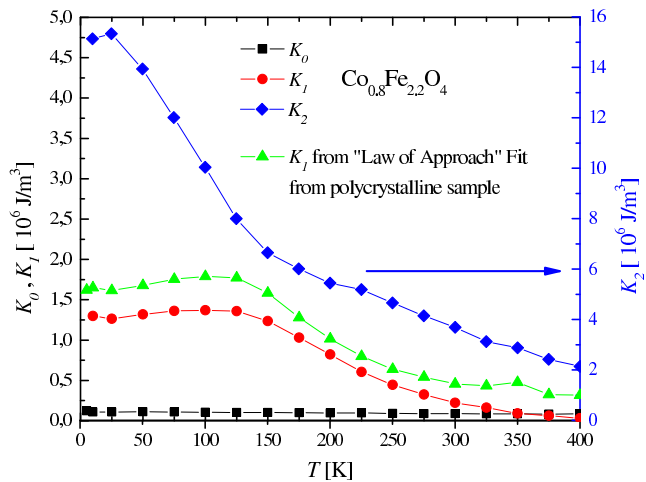}
\caption{\label{fig:EaConst}Anisotropy constants of Co$_{0.8}$Fe$_{2.2}$O$_{4}$ single crystal and one polycrystalline specimen as a function of temperature}
\end{figure}

The magnetic anisotropy was determined with the integral method $E_{A} = \int_{0}^{M_{S}} H dM$ along the measured crystallographic axis of the single crystal. Applying the formula $E_{A} = K_{0} + K_{1} (\alpha_{1}^{2}\alpha_{2}^{2} + \alpha_{2}^{2}\alpha_{3}^{2} + \alpha_{3}^{2}\alpha_{1}^{2}) + K_{2} (\alpha_{1}^{2}\alpha_{2}^{2}\alpha_{3}^{2})$, where the $\alpha_{i}$ are the direction cosines in polar coordinates, the anisotropy constants $K_{0}$, $K_{1}$ and $K_{2}$ were determined. For using the integral method it is pre-requisite to saturate the sample fully. It is reported that the saturation along the intermediate and hard axis is reached at around 18~T at $T = 4$~K \cite{guillot_high_1988}. Our measurements were only performed up to 9~T, therefore extrapolating the M(H) curves to saturation causes a rather large error ($\pm$ 20\% for $K_{1}$ and $\pm$ 30\% $K_{2}$), but they are still in the range of reported values of $K_{1}$ in literature.\\
In figure~\ref{fig:EaConst} the anisotropy constants $K_{0}$, $K_{1}$ and $K_{2}$ of the single crystal and $K_{1}$ of a polycrystalline sample are plotted against the temperature. For obtaining $K_{1}$ for the polycrystalline sample we used the law of approach to saturation. Measuring the polycrystalline sample the maximum magnetic field was only 9~T and the sample not fully saturated. Accordingly all reported values in literature using the law of approach to saturation are not giving the correct value of $K_{1}$, because they estimate only $K_{1}$ and not higher order anisotropy constants.\\
It is interesting to note that no values for $K_{2}$ are reported in literature. Above $T = 150$~K the value of $K_{2}$ is $\sim$~6 times higher than $K_{1}$ and below the FOMP the factor is even much higher, yielding in an increased anisotropy along the [111] and [110] axes.\\
\subsection{Magnetostriction}
\begin{figure}
\includegraphics{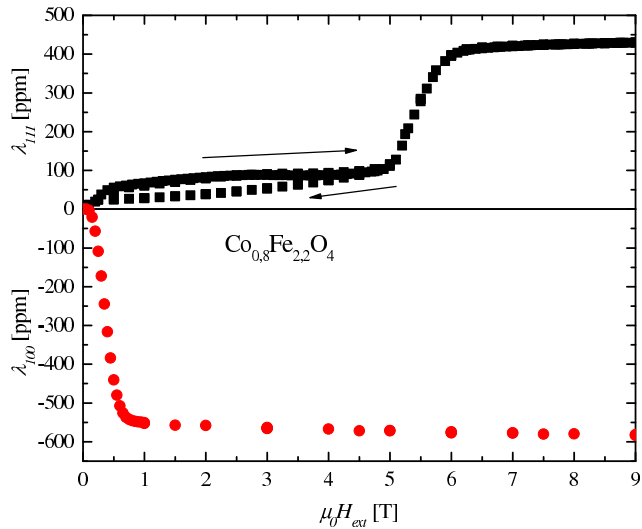}
\caption{\label{fig:MS4K}Magnetostriction measurements of Co$_{0.8}$Fe$_{2.2}$O$_{4}$ single crystal at $T = 4.2$K}
\end{figure}

We are presenting the first magnetostriction measurement at $T = 4.2$~K demonstrating also the effect of such a field induced transition in the magnetoelastic behaviour as shown in figure~\ref{fig:MS4K}. 
Due to hysteresis effects (remanence) the magnetostriction measurements at low temperatures are very difficult and need a special measuring procedure as will be published elsewhere. But when increasing the magnetic field above the critical field, a huge jump in the magnetostriction occurs. At higher fields due to rotational magnetization process no hysteresis effects are observed.\\
\section{Conclusions}
We have proved that the jump in the magnetization can be explained as an anisotropy driven transition which is called ``FOMP''. The transition is caused by a rotation of the magnetization vector jumping over an energy barrier. At low temperatures the second anisotropy constant $K_{2}$ is increasing which strengthens the [100] axis as easy axis and underlines the [110] as intermediate and the [111] as hard axis. These assumptions are also supported by geometric considerations explaining the values of the $M(H)$ curves as measured in the different crystallographic directions.
Additionally we present an accurate measurement of $\lambda_{111}$ at a temperature of 4.2~K showing also this critical transition in the magnetoelastic behaviour.\\
\begin{acknowledgments}
The financial support by the FWF under the NFN-project numbers S~10406 and S~10403 is gratefully acknowledged.
\end{acknowledgments}

%
\end{document}